\documentclass{article}
\usepackage{amssymb,amsmath,amsthm,latexsym,authblk,color}

\newtheorem{theorem}{Theorem}[section]
\newtheorem{lemma}[theorem]{Lemma}
\newtheorem{remark}[theorem]{Remark}
\newtheorem{definition}[theorem]{Definition}
\newtheorem{corollary}[theorem]{Corollary}
\newtheorem{proposition}[theorem]{Proposition}
\newtheorem{example}[theorem]{Example}

\numberwithin{equation}{section}
\numberwithin{table}{section}

\def\pf{\par{\bf Proof.}~ }
\def\C{{\mathbb C}}
\def\wh{\widehat} \def\wb{\overline}
\def \gdual{$G$-dual set}

\begin{document}

\title{Fourier Transforms and Bent Functions on Finite Abelian Group-Acted Sets}

\author[1]{Yun Fan} \author[2]{Bangteng Xu}
\affil[1]{School of Mathematics and Statistics,
 Central China Normal University, Wuhan 430079, China}
\affil[2]{Department of Mathematics and Statistics,
Eastern Kentucky University, Richmond, KY 40475, USA}

\insert\footins{\footnotesize
{\it Email addresses}:  yfan@mail.ccnu.edu.cn (Yun Fan),
bangteng.xu@eku.edu (Bangteng Xu)}

\maketitle

\begin{abstract}
Let $G$ be a finite abelian group acting faithfully on a finite set $X$.
As a natural generalization of the perfect nonlinearity of Boolean functions,
the $G$-bentness and $G$-perfect nonlinearity of functions on $X$ are studied
by Poinsot et al. \cite{P09, PH} via Fourier transforms of functions on $G$.
In this paper we introduce the so-called \gdual\ $\wh X$ of $X$,
which plays the role similar to the dual group $\wh G$ of $G$, and the
Fourier transforms of functions on $X$, a generalization of the Fourier
transforms of functions on finite abelian groups. Then we characterize the bent
functions on $X$ in terms of their own Fourier transforms on $\wh X$.
Bent (perfect nonlinear) functions on finite abelian groups and
$G$-bent ($G$-perfect nonlinear) functions on $X$ are treated in a uniform way
in this paper, and many known results in \cite{LSY, CD, P09, PH} are obtained as direct
consequences. Furthermore, we will prove that the bentness of a function on $X$
can be determined by its distance from the set of $G$-linear functions.
In order to explain the main results clearly, examples are also presented.

\end{abstract}

\section{Introduction}

Bent functions, perfect nonlinear functions, and their generalizations
have been studied in many papers.
The notion of a Boolean bent function
was introduced by Rothaus \cite{Rothaus}.
More than a decade ago, Logachev, Salnikov, and Yashchenko \cite{LSY}
generalized this concept to bent functions on finite abelian groups.
As a further generalization,
Poinsot \cite{P06} studied bent functions on finite nonabelian groups.
Recently, a closely related notion,
perfect nonlinear functions between finite abelian groups
as well as between arbitrary finite groups,
has been studied in quite a few papers;
for example, see \cite{CD, PP, Pott, S02, Xu1, Xu2}. These functions
have numerous applications in cryptography, coding theory, and other fields.
A critical tool in these studies is the Fourier transforms of functions
on finite groups.

The perfect nonlinearity of a function $f: G \to H$ between
finite abelian groups $G$ and $H$ is characterized by
its derivatives $f^\prime_\alpha:  G \to H, \,
x \mapsto f(\alpha x)f(x)^{-1}$, for all non-identity $\alpha \in G$.
By observing that, for any $\alpha\in G$,
the mapping $G \to G, \, x\mapsto \alpha x$ is just the regular action of $G$
on its base set $G$, Poinsot et al. \cite{P09, PH} generalized the concept of
the perfect nonlinearity to a function $g: X \to H$, where $X$ is a finite set
with an action of $G$ on it.  The derivatives of $g$ are
$g^\prime_\alpha: X \to H, \, x \mapsto f(\alpha x)f(x)^{-1}$, for any
$\alpha \in G$. In order to characterize the perfect nonlinearity
of functions from the set $X$ to $H$, the Fourier transforms of functions on
the group $G$ are used in \cite{P09, PH}.  For any $x \in X$,
let $g_x: G \to H$ be the function defined by
$g_x(\alpha) = g(\alpha x)$, for any $\alpha \in G$.
Let $\wh G, \wh H$ be the dual groups of $G$ and $H$, respectively,
and let $\zeta_0$ be the principal (irreducible) character of $H$.
Then $g$ is $G$-perfect nonlinear (cf. \cite[Theorems 2 and 3]{PH})
if and only if for any $\zeta \in \wh H \backslash \{ \zeta_0 \}$,
$$
\frac{1}{|X|} \sum_{x \in X} \big| \wh{\zeta{\circ}g_x}(\xi) \big|^2
= |G|, \quad \hbox{ for any } \xi \in \wh G,
$$
where $|X|$ and $|G|$ are the cardinalities of $X$ and $G$, respectively,
and $\wh{\zeta{\circ}g_x}$ is the Fourier transform of $\zeta{\circ}g_x$ on
$\wh G$. Poinsot \cite{P09} also defined the bentness of functions on $X$
in a similar way.
Let $\C$ be the complex field,  $T$ the unit circle in $\C$, and
 $f: X \to T$ a function.
For any $x \in X$, let $f_x$ be a function on $G$ defined by
$f_x(\alpha) := f(\alpha x)$, for any $\alpha \in G$.
Then using the Fourier transforms $\wh{f_x}$ of the functions~$f_x$ on~$G$,
the $G$-bentness of the function $f$ on $X$ is defined as follows:
$$
f \hbox{ is $G$-bent~ if~ } \frac{1}{|X|} \sum_{x \in X}
\big| \wh{f_x}(\xi) \big|^2 = |G|,~ \hbox{ for all } \xi \in \wh G.
$$
It is proved that
$f$ is $G$-bent if and only if for all non-identity element $\alpha \in G$,
the derivatives $f'_\alpha$ are balanced.

As the Fourier transform on finite groups is the key tool in the study of
bent and perfect nonlinear functions on finite groups, in this paper 
for a finite abelian group $G$ acting on a finite set $X$ ($X$ is called 
a {\em $G$-set}), we will develop the Fourier analysis on $X$, 
and use it as our tool to study the bentness
and perfect nonlinearity of functions on $X$.
By characterizing the bent
functions on $X$ in terms of their own Fourier transforms,
we are able to treat bent and perfect nonlinear functions
on finite abelian groups as well as
$G$-bent and $G$-perfect nonlinear functions on $X$ in a uniform way.

The set of functions from a $G$-set $X$ to $\C$, denoted by $\C^X$,
is a $\C G$-module, where $\C G$ is the group algebra of $G$ over $\C$.
$\C^X$ is also a unitary space with a natural $G$-invariant inner product.
For each irreducible character $\psi \in \wh G$,
the $G$-linear component of $\C^X$
with respect to $\psi$ is the $\C G$-submodule of $\C^X$
consisting of $\psi$-linear functions (see Definition \ref{defn-linear} below).
$\C^X$ can be decomposed into the orthogonal direct sum of
its $G$-linear components (see Proposition \ref{classical} below).
A key step is that by using this decomposition we obtain an orthogonal
basis $\wh X$ of $\C^X$ consisting of $G$-linear functions
such that $\wh X$ is closed under complex conjugation
(see Theorem \ref{hat X} below). Such a basis $\wh X$,
called a {\em \gdual} of $X$,
plays a role in $\C^X$ similar to~$\wh G$ in $\C^G$.
For any function $f \in \C^X$, we define the Fourier transform
$\wh f$ of $f$ as a function on $\wh X$ (see  Definition \ref{defn-four} below),
and define the bentness of $f$ in terms of $\wh{f}(\lambda)$
for all $\lambda \in \wh X$ (see Definition \ref{bent} below).
Our definitions of the Fourier transforms and
bentness of functions on $G$-sets are natural generalizations
of the Fourier transforms and bentness
of functions on finite abelian groups, respectively.

Then using the Fourier analysis on a $G$-set~$X$,
we study the characterizations of bent functions on $X$.
We will prove that (Theorem~\ref{bent deri-balance})
a function $f: X \to T$ is bent
if and only if the derivatives of~$f$ in all nontrivial directions are balanced.
Since the bentness and perfect nonlinearity of functions
on finite abelian groups and $G$-sets are treated in a uniform way,
many known results in \cite{LSY, CD, P09, PH} are obtained
as immediate consequences.
Furthermore, we will prove that (Theorem~\ref{bent max-dist})
$f \in T^X$ is a bent function if and only if the distance from $f$ to
the set $(\C^X)_G$ of $G$-linear functions reaches the best possible upper
bound of the distance between $(\C^X)_G$ and any function in $T^X$.
This result gives another geometric interpretation of
the importance of bent functions in cryptography.
The perfect nonlinearity of functions from $X$ to a finite abelian group $H$
is also characterized in terms of Fourier transforms of functions on $X$
(see Theorem \ref{thm-nonlinear}).
To explain the theory established in this paper,
several examples are also included.

The rest of the paper is organized as follows.
In Section \ref{sect-sets} we present the classical
decomposition of the $\C G$-module $\C^X$,
and prove the existence of the \gdual\ $\wh X$ of $X$.
Then in Section \ref{sect-four} we introduce the Fourier transforms
of functions in $\C^X$, and investigate their basic properties.
Section \ref{sect-bent} is devoted to the study of
the characterizations of bent functions on $X$.
Finally, $G$-perfect nonlinear functions are discussed
in Section \ref{sect-perf},
and explanatory examples are presented in Section \ref{sect-ex}.

\section{Complex functions and $G$-dual sets
\label{sect-sets}}

Throughout the paper, $G$ is always a finite abelian group of order $|G|=m$
with multiplicative operation,
$X$ a finite $G$-set with cardinality $|X|=n$,
$\C$ the complex field, and $T$ the unit circle in $\C$.
For any sets $R$ and $S$, by $R^S$ we denote the set of functions
from $S$ to $R$. Note that $\C^S$ is equipped with scalar multiplication,
function addition and function multiplication
such that it is a complex algebra. Also for $f\in\C^S$ by $\wb f$ we denote
the complex conjugation function, i.e. $\wb f(s)=\wb{f(s)}$ for $s\in S$,
where $\wb{f(s)}$ is the complex conjugate of $f(s)\in\C$. Furthermore,
$\C^X$ is a $\C G$-module (see below).
In this section we discuss the structure of $\C^X$,
and prove that it has a special basis $\wh X$, called the $G$-dual set,
which plays a similar role of $\wh G$ for the Fourier transforms.
Some of the results in this section are known for general
$\C G$-modules, but our treatment is different.

\subsection{Dual groups and Fourier transforms on groups}

An irreducible character of a finite abelian group $G$ is a homomorphism from $G$ to
the multiplicative group of non-zero complex numbers.
By $\wh G$ we denote the dual group of $G$, i.e. the group
consisting of irreducible characters of $G$.
For the fundamentals of representation theory of finite groups,
the reader is referred to~\cite{AB,Serre}.
We include some needed known facts here.

For any $\sigma\in\C^G$ we have a function $\wh\sigma\in\C^{\wh G}$,
called the {\em Fourier transform} of~$\sigma$, defined by
$\wh\sigma(\psi)=\sum_{\alpha\in G}\sigma(\alpha)\psi(\alpha)$,
for all $\psi\in\wh G$.
On the other hand, for any $\tau\in\C^{\wh G}$ we have
a function $\wh\tau\in\C^{G}$,
called the {\em Fourier inverse transform} of~$\tau$, defined as follows:
$\wh\tau(\alpha)=\frac{1}{m}\sum_{\psi\in\wh G}\tau(\wb\psi)\psi(\alpha)$
for all $\alpha\in G$.
Note that $\wb\psi\in\wh G$ for all $\psi\in\wh G$,
and $\wb\psi$ runs over $\wh G$ as $\psi$ runs over $\wh G$.
Also $\wb\psi(\alpha)=\psi(\alpha)^{-1}=\psi(\alpha^{-1})$,
for all $\alpha\in G$.

It is well-known that $\wh{\wh\sigma}=\sigma$ for all $\sigma\in\C^G$,
and $\wh{\wh\tau}=\tau$ for all $\tau\in\C^{\wh G}$.

\begin{remark}\label{rho}\rm
By $\rho$ we denote the regular character of $G$, i.e.
$$\rho(\alpha)=\begin{cases}m, & \alpha=1,\\ 0, &\alpha\ne 1; \end{cases}
  \qquad\forall~ \alpha\in G,
$$
where $1$ denotes the identity element of $G$.
It is known that $\rho=\sum_{\psi\in\wh G}\psi$.
Since $\wh G$ is a basis of the $m$-dimensional space $\C^G$,
any $\sigma\in\C^G$ is a linear combination of $\wh G$,
and the coefficients of the linear combination are uniquely determined by $\sigma$.
As $\sigma=\wh{\wh\sigma}$, by the Fourier inverse transform
 the linear combination of $\sigma$ is
$\sigma=\frac{1}{m}\sum_{\psi\in\wh G}\wh\sigma(\wb\psi)\psi$.
Note that $\sigma(\alpha)=0$ for all $\alpha\in G\backslash\{1\}$
if and only if $\sigma=\frac{\sigma(1)}{m}\rho$,
where $G\backslash\{1\}$ denotes the difference set
of $G$ removing the identity element~$1$. Thus,

(i)~ {\it $\sigma$ takes zero on $G\backslash\{1\}$ if and only if
$\wh\sigma$ is constant on $\wh G$.}

\noindent Since $G$ and $\wh G$
are dual to each other through the Fourier transforms,
by interchanging the roles of $\sigma$ and $\wh\sigma$ in (i), we further get

(ii)~ {\it $\sigma$ is constant on $G$ if and only if
$\wh\sigma$ takes zero on $\wh G\backslash\{1\}$,}

\noindent where by abuse of notation,
$1$ denotes the unity (principal) character: $1(\alpha)=1$
for all $\alpha\in G$.  No ambiguity of this notation can arise from the context.
\end{remark}

\subsection{$G$-linear functions and the classical decomposition}

As mentioned before, $X$ is a $G$-set with cardinality $|X|=n$. That is,
there is a map $G\times X \to X$, $(\alpha,x)\mapsto\alpha x$,
such that for all $x\in X$ we have
$(\alpha\beta)x=\alpha(\beta x)$ for all $\alpha,\beta\in G$,
and $1x=x$.
For the fundamentals of group actions, we refer the reader to~\cite{AB}.

If all the values of a function $f\in\C^X$ have length $1$, i.e.
$f\in T^X$, then we say that $f$ is a {\em unitary function}.

The complex space $\C^X$ is a  $\C G$-module
with the following $G$-action:
\begin{equation}\label{G act on C^X}
(\alpha f)(x)=f(\alpha^{-1}x),\qquad
 \forall~f\in\C^X~~\forall~\alpha\in G~~\forall~x\in X.
\end{equation}
In the literature, a $\C G$-module is also called a
complex $G$-space, see \cite[Ch 1]{Serre}. In this paper we will use both terms.
Furthermore, $\C^X$ is a unitary space with the following
inner product:
$$
\langle f,g\rangle=\sum_{x\in X}f(x)\wb g(x),\qquad\forall~~ f,g\in\C^X.
$$
This inner product is $G$-invariant in the following sense:
$\langle\alpha f,\,\alpha g\rangle=\langle f,\,g\rangle$,
or equivalently $\langle\alpha f,\,g\rangle=\langle f,\,\alpha^{-1}g\rangle$,
for all  $f,g\in\C^X$ and $\alpha\in G$.
The length (or norm) $|f|$ of any $f\in\C^X$ is then defined as
\begin{equation}
\label{eq-length}
|f|=\sqrt{\langle f,f\rangle}=\sqrt{\sum_{x\in X}f(x)\wb f(x)}\, ,
\end{equation}
and the distance between $f,g\in\C^X$ is defined as ${\rm d}(f,g)=|f-g|$.
Further, for any subsets $S_1,S_2\subseteq\C^X$ we can define
the distance between $S_1$ and~$S_2$ as follows:
\begin{equation}\label{dist sets}
{\rm d}(S_1,S_2)=\min
\big\{{\rm d}(f_1,f_2)\,\big|\, f_1\in S_1,\,f_2\in S_2\big\}.
\end{equation}

\begin{definition}
\label{defn-linear}
\rm (i)~
A function $f\in\C^X$ is said to be {\em $G$-linear}
if there is a
$\psi\in\wh G$ such that
$$
f(\alpha x)=\psi(\alpha)f(x),\qquad
 \forall~\alpha\in G~~\forall~x\in X,
$$
i.e. $\alpha f=\wb\psi(\alpha) f$ for all $\alpha\in G$.
More precisely, in that case we say that the function $f$ is {\em $\psi$-linear}.

(ii)~ For any $\psi\in\wh G$, the {\em $\psi$-component}
$(\C^X)_{\psi}$ of $\C^X$ is defined by
$$(\C^X)_{\psi}=\big\{f\,\big|\, f\in\C^X,~\mbox{$f$ is $\psi$-linear}\big\}.$$
The $\psi$-components for all $\psi\in\wh G$ are also simply called
{\em $G$-linear components} of~$\C^X$,
without mentioning the irreducible characters.

(iii)~ By $(\C^X)_{G}$ we denote the set of $G$-linear functions on $X$, i.e.
$$(\C^X)_{G}=\{f\mid f\in\C^X,\,\mbox{$f$ is $G$-linear}\}=
\bigcup_{\psi\in\wh G}(\C^X)_{\psi}.$$
\end{definition}

\begin{remark}\label{rem linear}\rm (i)~
Algebraically, a function $f \in \C^X$ is $G$-linear if and only if
$\tilde f:\C X\to\C$,
$ \sum_{x \in X} c_x x \mapsto \sum_{x \in X} c_x f(x)$
is a $\C G$-homomorphism, where $\C X$ is the permutation $\C G$-module
and $\C$ is viewed as a $\C G$-module with respect to some character $\psi$.
There is also a geometric interpretation of the $G$-linearity
by derivatives; see Lemma \ref{lem-linear} below.

(ii)~ Let $f$ be $\psi$-linear.
Since $\psi(\alpha)\ne 0$ for all $\alpha\in G$,
the zero-point set ${\rm Ann}(f)=\{x\in X\,|\,f(x)=0\}$ of
the function $f$ must be $G$-invariant.
Furthermore, if $\psi(\alpha)\ne 1$ for some $\alpha\in G$,
then $f(\alpha x)\ne f(x)$ (and hence $\alpha x\ne x$) for all $x\in\wb{\rm Ann}(f)$,
where $\wb{\rm Ann}(f)=\{x\in X\,|\,f(x)\ne 0\}$ denotes the
complement of ${\rm Ann}(f)$ in $X$.
\end{remark}

\begin{lemma}
\label{lem-component}
(i)~ For any $\psi \in \wh{G}$, $(\C^X)_\psi$ is a $G$-invariant subspace
$($i.e. $\C G$-submodule$)$ of $\C^X$.

(ii)~ For any $\psi, \varphi\in\wh G$ such that $\psi \ne \varphi$,
 we have the orthogonality:
$$
\big\langle(\C^X)_\psi,\,(\C^X)_\varphi\big\rangle=0.
$$
In particular, $(\C^X)_\psi\cap(\C^X)_\varphi=\{0\}$ if $\psi\ne\varphi$.
\end{lemma}

\pf (i) is straightforward. Now we prove (ii).
Since $\psi \ne \varphi$, there is an
$\alpha\in G$ such that $\psi(\alpha)\ne\varphi(\alpha)$.
So for any $f\in(\C^X)_\psi$ and $g\in(\C^X)_\varphi$,
$$
\psi(\alpha)\langle f, g\rangle=\langle\psi(\alpha) f, g\rangle
=\langle\alpha^{-1}f, g\rangle=\langle f,\alpha g\rangle
=\langle f,\wb\varphi(\alpha) g\rangle
=\varphi(\alpha)\langle f, g\rangle.
$$
Hence $\psi(\alpha)\ne \varphi(\alpha)$ implies that
$\langle f, g\rangle=0$.
This proves that $\big\langle(\C^X)_\psi,\,(\C^X)_\varphi\big\rangle=0$.
The rest of (ii) is clear. \qed

\begin{remark}\label{chi_f}\rm
The zero function $0$ is clearly $\psi$-linear for all $\psi\in\wh G$.
However, for a non-zero $G$-linear function $f$, Lemma \ref{lem-component}
implies that there is a unique $\psi\in\wh G$ such that $f$ is $\psi$-linear.
In other words, the union $(\C^X)_G=\bigcup\limits_{\psi\in\wh G}(\C^X)_\psi$
is a disjoint union, in the sense that only the zero function is
in the intersections.
\end{remark}

For $\sigma\in\C^G$ and $f\in\C^X$, we define a function $\sigma{*}f\in\C^X$,
called the {\em convolution product} of $\sigma$ and $f$, as follows:
$$
(\sigma{*}f)(x)=\sum_{\alpha\in G}\sigma(\alpha)f(\alpha^{-1}x),
\qquad\forall~~x\in X.
$$
Specifically, taking $X=G$ to be the regular $G$-set, then
for $\sigma,\tau\in\C^G$ the above formula gives the usual convolution product
 $\tau{*}\sigma\in\C^G$ of functions on the group:
$$
 (\tau{*}\sigma)(\alpha)=\sum_{\beta\in G}
   \tau(\beta)\sigma(\beta^{-1}\alpha),\qquad\forall~~\alpha\in G.
$$

\begin{lemma}
\label{lem-conv}
(i)~ The following map is bilinear:
$$
 \C^G\times\C^X~\longrightarrow~\C^X,~~~ (\sigma,\,f)\longmapsto\sigma{*}f.
$$

(ii)~ $(\tau{*}\sigma){*}f=\tau{*}(\sigma{*}f)$,\quad$\forall$~
  $\tau,\sigma\in\C^G,~f\in\C^X$.

(iii)~ $\wb{\sigma{*}f} = \wb{\sigma}*\wb{f}$,\quad$\forall$~
  $\sigma\in\C^G,~f\in\C^X$.
\end{lemma}

\pf The lemma is true by straightforward computations.\qed

Any $G$-space is decomposed into a direct sum with
each summand associated with exactly one irreducible character;
such a decomposition is called the {\em classical decomposition}
(see \cite[\S2.6]{Serre}). The following proposition shows a practical way to
compute the classical decomposition of $\C^X$.

\begin{proposition}\label{classical}
(i)~ For any $\psi\in\wh G$ and $f\in\C^X$, we have
$\psi{*}f\in(\C^X)_\psi$.

(ii)~ For any $\psi\in\wh G$ and $f\in\C^X$, $f \in (\C^X)_\psi$ if and
only if $ f = \frac{1}{m} \psi{*}f$.

(iii)~ For any $f\in C^X$, we have
$\displaystyle f=\frac{1}{m}\sum\limits_{\psi\in\wh G}\psi{*}f$.

(iv)~
We have the orthogonal direct sum:
$\C^X=\bigoplus\limits_{\psi\in\wh G}(\C^X)_\psi$.
\end{proposition}

\pf (i).~ For any $\alpha\in G$ and $x\in X$, since $\alpha\beta$ runs over $G$
as $\beta$ runs over $G$, we have
\begin{eqnarray*}
(\psi{*}f)(\alpha x)&=&\sum_{\beta\in G}
\psi(\alpha\beta)f((\alpha\beta)^{-1}\alpha x)
=\sum_{\beta\in G}\psi(\alpha)\psi(\beta)f(\beta^{-1}x)\\
&=&\psi(\alpha)\cdot\sum_{\beta\in G}\psi(\beta)f(\beta^{-1}x)
=\psi(\alpha)\cdot(\psi{*}f)(x).
\end{eqnarray*}

(ii).~ If $f = \frac{1}{m} \psi{*}f$, then $f \in (\C^X)_\psi$
by (i). If $f \in (\C^X)_\psi$, then for any $x \in X$,
$$
\frac{1}{m} (\psi{*}f) (x)
= \frac{1}{m} \sum_{\alpha \in G} \psi(\alpha)f(\alpha^{-1}x)
= \frac{1}{m} \sum_{\alpha \in G} \psi(\alpha) \wb{\psi}(\alpha)f(x) = f(x).
$$

(iii).~ Since the regular character $\rho$ satisfies that
$\rho(\alpha)=\begin{cases}0, &\alpha\ne 1;\\ m,&\alpha=1;\end{cases}$,
we have $\frac{1}{m}\rho{*}f=f$, for all $f\in C^X$.
Furthermore, since $\rho=\sum_{\psi\in\hat G}\psi$, we get that
$$
f=\frac{1}{m}\rho{*}f=\frac{1}{m}\Big(\sum_{\psi\in\wh G}\psi\Big){*}f
=\frac{1}{m}\sum_{\psi\in\wh G}\psi{*}f.
$$

(iv) follows directly from (i), (iii), and Lemma \ref{lem-component}.
\qed

\begin{definition}\rm
(i)~ For any $f\in C^X$ and $\psi\in\wh G$,
the {\em $\psi$-component} of $f$ is $f_\psi=\frac{1}{m}\psi{*}f$,
and the {\em classical decomposition} of $f$ is $f=\sum_{\psi\in\wh G}f_\psi$.

(ii)~ The orthogonal direct sum
$\C^X=\bigoplus\limits_{\psi\in\wh G}(\C^X)_\psi$
is called the {\em classical decomposition} of the $G$-space $\C^X$;
cf. \cite[\S2.6]{Serre}.
\end{definition}

\subsection{The \gdual\ $\wh X$}

For an unitary space $V$ and a basis $u_1,\cdots,u_n$ of $V$, if
there is a non-zero $\wb n\in\C$ such that
$\langle u_i,u_j\rangle=\begin{cases}0, & i\ne j;\\\wb n, & i=j;\end{cases}$,
then we say that $u_1,\cdots,u_n$ is an {\em $\bar n$-normal
orthogonal basis} of the unitary space.

\begin{definition}\label{def standard}\rm
A basis $\wh X$ of the unitary $G$-space $\C^X$
is called a {\em \gdual\ of $X$} if the following three conditions are satisfied:

(i)~ $\wh X$ is an $n$-normal orthogonal basis;

(ii)~ any $\lambda\in\wh X$ is $G$-linear;

(iii)~ $\wh X$ is closed under complex conjugation, i.e.
$\wb\lambda\in\wh X$ for all $\lambda\in\wh X$.
\end{definition}

\begin{theorem}\label{hat X}
For any $G$-set $X$, there exists a \gdual\ $\wh X$.
\end{theorem}

\pf Since $\wb\psi\in\wh G$ for any $\psi\in\wh G$, it follows from Lemma
\ref{lem-conv}(iii) and Proposition \ref{classical}(ii) that
for any $f\in(\C^X)_\psi$, $\wb f\in(\C^X)_{\wb\psi}$. That is,
$$
\wb{(\C^X)}_{\psi}=(\C^X)_{\wb\psi},\qquad \forall~~ \psi\in\wh G,
$$
where $\wb{(\C^X)}_{\psi}=\big\{\wb f\,\big|\,f\in (\C^X)_{\psi}\big\}$.
For any $\psi\in\wh G$, it is known that there is an $n$-normal orthogonal basis
$(\wh X)_\psi$ for the $\psi$-component $(\C^X)_\psi$ of $\C^X$.
Hence,
$\wb{(\wh X)}_\psi=\big\{\wb\lambda\,\big|\,\lambda\in(\wh X)_\psi\big\}$
is also an $n$-normal orthogonal basis
of the $\wb\psi$-component $(\C^X)_{\wb\psi}$. Thus, if $\psi \ne \wb\psi$,
then $(\wh X)_\psi \cup \wb{(\wh X)}_\psi$ is an $n$-normal orthogonal basis of
$(\C^X)_\psi \oplus (\C^X)_{\wb\psi}$.

In the following we prove that if $\psi=\wb\psi$, then there is an
$n$-normal orthogonal basis $(\wh X)_\psi$ of $(\C^X)_\psi$ such that for any
$\lambda \in (\wh X)_\psi$, $\lambda = \wb\lambda$. Let $f \in (\C^X)_\psi$
such that $f \ne 0$. Then at least one of $f + \wb f$ and $\sqrt{-1}(f - \wb f)$
is not zero. Thus, $\wb{(\C^X)}_\psi = (\C^X)_\psi$ implies that
there is a  $\lambda_1 \in (\C^X)_\psi$ such that $\lambda_1 \ne 0$,
and $\lambda_1 = \wb\lambda_1$. We may also assume that
$\langle \lambda_1, \lambda_1\rangle = n$.
Note that $(\C^X)_\psi = \C \lambda_1 \oplus (\C \lambda_1)^\perp$.
Also for any $f \in (\C \lambda_1)^\perp$,
it follows from $\lambda_1 = \wb\lambda_1$
that $\wb{f} \in (\C \lambda_1)^\perp$. Hence,
if $(\C \lambda_1)^\perp \ne \{ 0 \}$, then as above,
there is $\lambda_2 \in (\C \lambda_1)^\perp$ such that
$\lambda_2 = \wb\lambda_2$, $\langle \lambda_2, \lambda_2 \rangle = n$,
and $(\C \lambda_1)^\perp = \C \lambda_2 \oplus
 (\C \lambda_1\oplus\C \lambda_2)^\perp$.
Continuing this process, we see that $\lambda_1, \lambda_2, \cdots$
form an $n$-normal orthogonal basis of $(\C^X)_\psi$.

Therefore, the orthogonal direct sum
$\C^X=\bigoplus_{\psi\in\wh G}(\C^X)_\psi$ implies that
the union $\wh X$ of the $n$-normal orthogonal bases of the
$G$-linear components of $\C^X$ chosen in the above two paragraphs is a
basis of $\C^X$ that satisfies the conditions (i), (ii) and (iii)
of Definition~\ref{def standard}.
\qed

\begin{remark}\label{rem standard}\rm
(i)~ If $\wh X$ is a \gdual\ of $X$, then
$\wh Y=\{\varepsilon\lambda\,|\,\lambda\in\wh X,\,\varepsilon\in T\}$
is also a \gdual\ of $X$. We call $\wh Y$ a rescaling of $\wh X$ by $T$.

(ii)~ If $X$ is a transitive $G$-set, then every $G$-linear component
$(\C^X)_\psi$ of $\C^X$ is $1$-dimensional, and
hence $(\wh X)_{\psi}$ consists of exactly one function of length $\sqrt n$.
Thus, $X$ has a unique \gdual\ $\wh X$ up to rescaling by $T$.

(iii)~ In particular, if $X=G$ is the regular $G$-set, then $X$ has a
unique \gdual\ up to rescaling by $T$. Usually,
the dual group $\wh G$ is chosen as $\wh X$.

(iv)~ However, if the number of the $G$-orbits of $X$ is greater that $1$,
then the \gdual~$\wh X$ is not unique up to rescaling by $T$.
The proof  of Theorem~\ref{hat X} provides a way to chose a \gdual.
Later we will show another way to get a \gdual
(see Examples~\ref{K4 no} and \ref{K4 yes} below).
\end{remark}

From now on, for the  $G$-set $X$ we fix a \gdual\ $\wh X$.
Then we have the disjoint union
$$
\wh X=\bigcup_{\psi\in\wh G}(\wh X)_\psi,
$$
where $(\wh X)_\psi$ is an $n$-normal orthogonal basis of $(\C^X)_\psi$,
and $\wb{(\wh X)_\psi}=(\wh X)_{\wb\psi}$.
Thus, the $\psi$-component of $\C^X$
is
\begin{equation}\label{dec hat X}
(\C^X)_\psi=\bigoplus_{\lambda\in(\wh X)_\psi}\C\lambda,\qquad
\forall~~\psi\in\wh G.
\end{equation}
Note that some subsets $(\wh X)_\psi$ may be empty
(correspondingly, some component $(\C^X)_\psi$ may be zero).

Let $\wh X=\{\lambda_1,\cdots,\lambda_n\}$ and $X=\{x_1,\cdots,x_n\}$.
Then we have an $n \times n$ matrix
$\Lambda=\big(\lambda_i(x_j)\big)_{1\le i, j\le n}$.
The $n$-normal orthogonality of $\wh X$ implies that
$\Lambda\cdot\wb\Lambda^T=nI$, where $I$ is the identity matrix and
$\wb\Lambda^T$ is the conjugate transpose of $\Lambda$.
Hence we also have $\wb\Lambda^T\cdot\Lambda=nI$.
Thus, we have the following

\begin{lemma}
\label{lem-or}
{\rm (Orthogonality Relations)}
The following hold:
\begin{equation}\label{1-ortho}
\sum_{x\in X}\lambda(x)\wb\mu(x)
 =\begin{cases}n, &\lambda=\mu;\\ 0, & \lambda\ne\mu;\end{cases}
 \qquad\forall~~\lambda,~\mu\in\wh X.
\end{equation}
\begin{equation}\label{2-ortho}
\sum_{\lambda\in\wh X}\lambda(x)\wb\lambda(y)
 =\begin{cases}n, & x=y;\\ 0, & x\ne y;\end{cases}
 \qquad\forall~~x,~y\in X.
\end{equation}
\end{lemma}

\section{Fourier transforms of functions on $G$-sets}
\label{sect-four}

Given a \gdual\ $\wh X$ of the $G$-set $X$, in this section we
define the Fourier transform of $f \in \C^X$ on $\wh X$, and discuss its
basic properties. We will need to consider the space $\C^{\wh X}$ of complex
functions on $\wh X$, which is also a unitary space with the inner product
$$
\langle g,h \rangle=\sum_{\lambda\in\wh X}g(\lambda)\bar h(\lambda),
\qquad\forall~~ g,h\in\C^{\wh X}.
$$

For any $\sigma \in \C^G$, the Fourier transform $\wh\sigma$
of $\sigma$ at any $\psi \in \wh G$ is
$\wh{\sigma}(\psi) = \sum_{\alpha \in G} \sigma(\alpha) \psi(\alpha)$.
The next definition generalizes this notion to the functions on $G$-sets.

\begin{definition}
\label{defn-four}
\rm
For any $f\in\C^X$, the {\em Fourier transform} of $f$,
 $\wh f\in\C^{\wh X}$, is defined as
$$
\wh f(\lambda)=\sum_{x\in X}f(x)\lambda(x),\qquad\forall~~\lambda\in\wh X.
$$
For any $g\in\C^{\wh X}$, the {\em Fourier inversion} of $g$,
 $\wh g\in\C^{X}$, is defined as
$$
\wh g(x)=\frac{1}{n}\sum_{\lambda\in\wh X}g(\lambda)\wb\lambda(x),
\qquad\forall~~x\in X.
$$
\end{definition}

It is clear that if $X = G$ is the regular $G$-set,
then the Fourier transform of $f \in \C^X$ in the above definition
is exactly the Fourier transform of functions on~$G$.

\begin{remark}
\label{re-four}
\rm
For $x\in X$ we have the characteristic function
${\bf 1}_x$ (i.e. ${\bf 1}_x(y)=0$, if $y\ne x$, and ${\bf 1}_x(x)=1$),
whose Fourier transform is $\wh{\bf 1}_x(\lambda)=\lambda(x)$, for any
$\lambda\in\wh X$. Thus, we can rewrite the definitions of $\wh f$ and
$\wh g$ in Definition \ref{defn-four} as follows:
$$
\wh f(\lambda)=\langle f,\wb\lambda\rangle,\qquad
\forall~f\in\C^X,~~\forall~\lambda\in\wh X;
$$
$$
\wh g(x)=\frac{1}{n}\langle g,\wh{\bf 1}_x\rangle,\qquad
 \forall~g\in\C^{\wh X},~~\forall~x\in X.
$$
\end{remark}

\begin{lemma}~
\label{lem-hat-hat}
$\wh{\wh f}=f,\quad\forall~f\in\C^X$, \ \ and \ \
 $\wh{\wh g}=g,\quad\forall~g\in\C^{\wh X}.$
\end{lemma}

\pf For any $x\in X$ we have
$$
\wh{\wh f}(x) 
=\frac{1}{n}\sum_{\lambda\in\wh X}
\Big(\sum_{y\in X}f(y)\lambda(y)\Big)\wb\lambda(x)
=\sum_{y\in X}f(y)\cdot\frac{1}{n}
\sum_{\lambda\in\wh X}\lambda(y)\wb\lambda(x).
$$
By the second orthogonality relation~(\ref{2-ortho}),
we have $\wh{\wh f}(x)=f(x)$ for all $x\in X$.
Similarly, by the first orthogonality relation~(\ref{1-ortho}), we have
$\wh{\wh g}(\lambda)=g(\lambda)$ for all $\lambda\in\wh X$.
\qed

For any $\lambda\in\wh X$, there is a unique irreducible character $\psi_\lambda$
of $G$ such that $\lambda\in(\C^X)_{\psi_\lambda}$ by Remark~\ref{chi_f}.
Also for any $g\in\C^{\wh X}$, the length of $g$ is $|g| = \sqrt{\langle g,
g \rangle}$.

\begin{lemma}\label{hat*}
Let $\sigma\in\C^G$ and $f\in\C^X$. Then the following hold.

(i)~ $\wh{\sigma{*}f}(\lambda)=\wh\sigma(\psi_\lambda)\wh f(\lambda)$
  ~for all $\lambda\in\wh X$.

(ii)~ If $\sigma\in\wh G$,
then $\wh{\frac{1}{m}\sigma{*}f}(\lambda)
=\begin{cases}\wh f(\lambda), &\lambda\in(\wh X)_{\wb\sigma};\\
 0,&\lambda\notin(\wh X)_{\wb\sigma}.\end{cases}$

(iii)~ $|\wh{f_\psi}|^2=|\wh{\frac{1}{m}\psi{*}f}|^2=\sum\limits_{
\lambda\in(\wh X)_{\psi}}|\wh f(\wb\lambda)|^2$, for all $\psi\in\wh G$.
\end{lemma}

\pf (i).~ Since $\lambda(\alpha^{-1}x)=\psi_\lambda(\alpha^{-1})\lambda(x)$
for $\alpha\in G$ and $x\in X$, we have
\begin{eqnarray*}
\wh{\sigma{*}f}(\lambda)&=&\sum_{x\in X}(\sigma{*}f)(x)\lambda(x)
=\sum_{x\in X}\sum_{\alpha\in G}\sigma(\alpha)f(\alpha^{-1}x)\lambda(x)\\
&=&\sum_{\alpha\in G}\sigma(\alpha)
\psi_\lambda(\alpha)\sum_{x\in X}f(\alpha^{-1}x)\lambda(\alpha^{-1}x)\\
&=&\wh\sigma(\psi_\lambda)\wh f(\lambda).
\end{eqnarray*}

(ii).~ If $\sigma\in\wh G$, then by Remark \ref{re-four},
$\wh\sigma(\psi_\lambda)=\langle\sigma,\wb\psi_\lambda\rangle
=\begin{cases}m, &\sigma=\wb\psi_\lambda;\\
 0,&\sigma \ne \wb\psi_\lambda.\end{cases}$
So (ii) follows from (i).

(iii).~ $|\wh{f_\psi}|^2=|\wh{\frac{1}{m}\psi{*}f}|^2
=\sum_{\lambda\in\wh X}|\wh{\frac{1}{m}\psi{*}f}(\lambda)|^2$.
Since $\wh X=\bigcup_{\psi\in\wh G}(\wh X)_{\psi}$,
by (ii) we get $|\wh{f_\psi}|^2
=\sum_{\lambda\in(\wh X)_{\wb\psi}}|\wh f(\lambda)|^2
=\sum_{\lambda\in(\wh X)_{\psi}}|\wh f(\wb\lambda)|^2$.
\qed

The following is an easy but useful fact.

\begin{lemma}\label{linear comb}
Any function $f\in\C^X$ is a unique linear combination of $\wh X$ as follows:
$$f=\frac{1}{n}\sum\limits_{\lambda\in\wh X}\wh f(\wb\lambda)\lambda.$$
Hence, for the classical decomposition
$f=\sum\limits_{\psi\in\wh G}f_\psi$,
the $\psi$-component
$$f_\psi=\frac{1}{m}\psi{*}f=\frac{1}{n}\sum\limits_{\lambda\in(\wh X)_\psi}
  \wh f(\wb\lambda)\lambda.
$$
\end{lemma}

\pf ~ For all $x\in X$, Lemma \ref{lem-hat-hat} implies that
$$
f(x)=\wh{\wh f}(x)
=\frac{1}{n}\sum_{\lambda\in\wh X}\wh f(\lambda)\wb\lambda(x)
=\frac{1}{n}\sum_{\lambda\in\wh X}\wh f(\wb\lambda)\lambda(x).
$$
Since $\displaystyle \frac{1}{n}\sum\limits_{\lambda\in(\wh X)_\psi}
  \wh f(\wb\lambda)\lambda \in (\C^X)_\psi$, Proposition \ref{classical}
and \eqref{dec hat X}  imply that the $\psi$-component of $f$ is
$f_\psi=\frac{1}{n}\sum_{\lambda\in(\wh X)_\psi}
  \wh f(\wb\lambda)\lambda$. \qed

\begin{lemma}\label{<f,g>}
Let $f,g\in\C^X$ and $\alpha\in G$. Then
$$
 \langle{\alpha^{-1}}f,g\rangle
  =\frac{1}{n}\sum_{\psi\in\wh G}\psi(\alpha)
   \sum_{\lambda\in(\wh X)_{\psi}}\wh f(\wb\lambda)\wb{\wh g}(\wb\lambda).
$$
In particular
$
 \langle f,g\rangle=\frac{1}{n}\langle\wh f,\wh g\rangle.
$
\end{lemma}

\pf Recall that $(\alpha^{-1}f)(x)=f(\alpha x)$
(cf. Eqn~(\ref{G act on C^X})). So by Lemma~\ref{linear comb}, we have
\begin{eqnarray*}
\langle \alpha^{-1}f,g\rangle=
\sum_{x\in X}f(\alpha x)\wb g(x)
=\sum_{x\in X}\frac{1}{n}\sum_{\lambda\in\wh X}
  \wh f(\wb\lambda)\lambda(\alpha x)
   \frac{1}{n}\sum_{\mu\in\wh X}\wb{{\wh g}(\wb\mu)\mu(x)}.
\end{eqnarray*}
Note that $\wh X$ is the disjoint union
$\wh X=\bigcup_{\psi\in\wh G}(\wh X)_\psi$,
and for $\lambda\in(\wh X)_\psi$
we have $\lambda(\alpha x)=\psi(\alpha)\lambda(x)$. So
\begin{eqnarray*}
\langle \alpha^{-1}f,g\rangle&=&\frac{1}{n^2}\sum_{x\in X}
\sum_{\psi\in\wh G}\sum_{\lambda\in(\wh X)_\psi}
  \wh f(\wb\lambda)\psi(\alpha)\lambda(x)
 \sum_{\mu\in\wh X}\wb{\wh g}(\wb\mu)\wb\mu(x)\\
&=&\frac{1}{n^2}\sum_{\psi\in\wh G}\psi(\alpha)\sum_{\lambda\in(\wh X)_\psi}
 \sum_{\mu\in\wh X} \wh f(\wb\lambda)\wb{\wh g}(\wb\mu)
 \sum_{x\in X} \lambda(x)\wb\mu(x).
\end{eqnarray*}
By the first orthogonality relation~(\ref{1-ortho}), we get that
$$\langle \alpha^{-1}f,g\rangle=\frac{1}{n}\sum_{\psi\in\wh G}
\psi(\alpha)\sum_{\lambda\in(\wh X)_\psi}
 \wh f(\wb\lambda)\wb{\wh g}(\wb\lambda).
$$
Taking $\alpha=1$ in the above formula, we have
$\langle f,g\rangle=\frac{1}{n}\langle\wh f,\wh g\rangle$.
\qed

The next corollary is immediate from Lemma \ref{<f,g>}.

\begin{corollary}\label{cor13}
If $f\in T^X$, then
$\langle f,f\rangle=n$ and $\langle \wh f,\wh f\rangle=n^2$.
\end{corollary}

For any $\lambda\in\wh X$, $\wh\lambda \in \C^{\wh X}$, and $\wh\lambda (\mu) =
\begin{cases}0, & \mu \ne \wb\lambda;\\ n, & \mu = \wb\lambda.\end{cases}$
So $\{\wh\lambda\mid\lambda\in\wh X\}$ is an $n^2$-normal orthogonal
basis of $\C^{\wh X}$.

\section{Bent functions on $G$-sets}
\label{sect-bent}

In this section we define the bentness of functions on the $G$-set $X$, and
study its characterizations.
For a finite abelian group $G$, a unitary function $f: G \to T$ is called a
{\em bent function} (cf. \cite{LSY}) if for any $\psi \in \wh G$,
$|\wh{f}(\psi)|^2 = |G|$.
The next definition generalizes this notion to unitary functions on $G$-sets.

\begin{definition}\label{bent}\rm
A unitary function $f:X\to T$ on the $G$-set $X$ is called a
{\em bent function} if
$$
\sum\limits_{\lambda\in(\wh X)_{\psi}} \big | \wh f(\lambda) \big |^2
=\frac{|X|^2}{|G|}, \quad \hbox{for all } \psi\in\wh G.
$$
\end{definition}

If $X = G$ is the regular $G$-set, then for any $\psi\in\wh G$,
$(\wh X)_{\psi} = \{ \psi \}$, and the above definition of
a bent function on the $G$-set $X$ is exactly the same as
the definition of a bent function on $G$.
The bentness of functions on $G$-sets
are also defined in \cite[Definition 6]{P09},
and called {\em $G$-bent functions}. But the definition in \cite{P09}
is different; it uses the Fourier transforms of functions on $G$.
However, we will show that the definition in \cite{P09} is
equivalent to Definition \ref{bent} (see Corollary \ref{cor-bent} below).

Although the bent function is defined by the use of
$\lambda \in\wh X$, the next lemma says that the bentness of a function
on $X$ is independent of the choice of $\wh X$.
Recall that the length of a function is defined by Eqn \eqref{eq-length}.

\begin{lemma}
\label{lem-bent}
For a unitary function $f:X\to T$, the following are equivalent.

(i)~ $f$ is a bent function.

(ii)~ For any $\psi, \varphi \in\wh G$, $|\wh f_\psi| = |\wh f_\varphi|$.

(iii)~ For any $\psi, \varphi \in\wh G$, $|f_\psi| = |f_\varphi|$.
\end{lemma}

\pf
By Lemma \ref{hat*}, (i) implies (ii). Assume (ii).
From Lemma \ref{hat*} and Corollary~\ref{cor13} we see that
$$\sum_{\psi\in\wh G}|\wh{f_\psi}|^2
=\sum_{\psi\in\wh G}\sum_{\lambda\in(\wh X)_{\wb\psi}}|\wh f(\lambda)|^2
=\langle\wh f,\wh f\rangle=n^2.
$$
Hence, for any $\psi \in\wh G$,
$\sum_{\lambda\in(\wh X)_{\psi}} \big | \wh f(\lambda) \big |^2 =
|\wh{f_\psi}|^2 = n^2/m$, and (i) holds.

(ii) and (iii) are equivalent by Lemma \ref{<f,g>}. \qed

The zero-point set ${\rm Ann}(f)$ of $f\in\C^X$ and its complement
$\wb{\rm Ann}(f)$ in $X$ are introduced in Remark~\ref{rem linear}.
Note that $f$ is non-zero if and only if $\wb{\rm Ann}(f)\ne\emptyset$.

\begin{definition}\rm
If $f\in\C^X$ is a non-zero function and ${\rm Ann}(f)$ is $G$-invariant,
then $f$ is said to be {\em differentiable}.
For any differentiable function $f\in\C^X$  we define a function
$f'_{\alpha}$ on $\wb{\rm Ann}(f)$ as follows:
$$
 f'_{\alpha}(x)=f(\alpha x)f(x)^{-1},\qquad \forall~~ x\in\wb{\rm Ann}(f).
$$
$f'_{\alpha}$ is called the {\em derivative} of $f$ in direction $\alpha$.
\end{definition}

Any unitary function $f\in T^X$ is differentiable and $f'_\alpha\in T^X$.
By Remark \ref{rem linear}(ii),
any non-zero $G$-linear function is differentiable.
The following lemma is a geometric explanation of the $G$-linearity of a function
by its derivative.

\begin{lemma}\label{lem-linear} Let $f\in\C^X$ be differentiable. Then
$f'_{\alpha}$ is a constant function on $\wb{\rm Ann}(f)$
for any $\alpha\in G$
if and only if $f$ is $G$-linear, i.e.
there is a unique character $\psi\in\wh G$ such that $f\in(\C^X)_\psi$.
\end{lemma}

\pf
It is clear that if $f$ is $\psi$-linear for $\psi\in\wh G$,
then for any $\alpha\in G$,
$f'_{\alpha}(x)=\psi(\alpha)$ for $x\in\wb{\rm Ann}(f)$
is a constant function. Now assume that
for any $\alpha\in G$, $f'_{\alpha}(x)=\psi_f(\alpha)$,
for all $x\in\wb{\rm Ann}(f)$.
Then for any $\alpha,\beta\in G$ we have
\begin{eqnarray*}
\psi_f({\alpha\beta})&=&f'_{\alpha\beta}(x)=f(\alpha\beta x)f(x)^{-1}\\
&=&f(\alpha(\beta x))f(\beta x)^{-1}\cdot f(\beta x)f(x)^{-1}\\
&=&f'_{\alpha}(\beta x)\cdot f'_{\beta}(x)=\psi_f(\alpha)\psi_f(\beta).
\end{eqnarray*}
So $\psi_f$ is an irreducible character of $G$, and
$$
f(\alpha x)=\psi_f(\alpha)f(x),\qquad\forall~~x\in X.\eqno\qed
$$

Unitary functions far away from $G$-linear functions on $X$ are more useful and
interesting in cryptography.
So by Lemma \ref{lem-linear} we want to investigate those unitary functions
whose derivatives in all nontrivial directions
are far away from constant functions.
As for unitary functions on finite groups, a unitary function $h:X\to T$
is said to be {\em balanced} if $\sum_{x\in X}h(x)=0$.

\begin{definition}\rm
A unitary function $f: X\to T$ is said to have
{\em totally balanced derivatives} if
$$
 \sum_{x\in X}f'_\alpha(x)=0,\qquad\forall~~\alpha\in G\backslash\{1\}.
$$
\end{definition}

\medskip
Now we are ready to present the characterizations of bent functions on $G$-sets.

\begin{theorem}\label{bent deri-balance}
A unitary function $f\in T^X$ is bent
if and only if $f$ has totally balanced derivatives.
\end{theorem}

\pf Note that
$\sum\limits_{x\in X}f'_\alpha(x)=\sum_{x\in X}f(\alpha x)\wb f(x)=
\langle\alpha^{-1}f,f\rangle$. So by Lemma~\ref{<f,g>} we have
\begin{eqnarray*}
\sum_{x\in X}f'_\alpha(x)
&=&\frac{1}{n}\sum_{\psi\in\wh G}\psi(\alpha)
   \sum_{\lambda\in(\wh X)_{\psi}}\wh f(\wb\lambda)\wb{\wh f}(\wb\lambda)\\
&=&\frac{1}{n}\sum_{\psi\in\wh G}
 \Big(\sum_{\lambda\in(\wh X)_\psi}|\wh f(\wb\lambda)|^2\Big)\psi(\alpha).
\end{eqnarray*}
Thus, Lemma~\ref{hat*}(iii) implies that
\begin{equation}\label{f'}
\sum_{x\in X}f'_\alpha(x)=\frac{1}{n}\sum_{\psi\in\wh G}
|\wh{f_\psi}|^2\psi(\alpha).
\end{equation}

If $f$ has totally balanced derivatives, i.e.
$\sum_{x\in X}f'_\alpha(x)=0$ for all $\alpha\in G\backslash\{1\}$,
then Eqn~\eqref{f'} implies that
the function $\sum_{\psi\in\wh G}|\wh{f_\psi}|^2\psi$ on~$G$
takes zero on $G\backslash\{1\}$,
and hence it must be a multiple of the regular character
$\rho=\sum_{\psi\in\wh G}\psi$ of $G$, cf. Remark~\ref{rho}.
Thus, for any $\psi,\varphi\in\wh G$ we have
$|\wh{f_\psi}|^2=|\wh{f_\varphi}|^2$,
and $f$ is bent by Lemma \ref{lem-bent}.

Conversely, if $f$ is bent, i.e.
$|\wh{f_\psi}|^2=\frac{n^2}{m}$
for all $\psi\in\wh G$, then by Eqn~(\ref{f'}) we have
$$
\sum_{x\in X}f'_\alpha(x)=
\frac{1}{n}\sum_{\psi\in \wh G}\frac{n^2}{m}\psi(\alpha)
=\frac{n}{m}\sum_{\psi\in\wh G}\psi(\alpha)=\frac{n}{m}\rho(\alpha).
$$
So for all $\alpha\in G\backslash\{1\}$,
$\sum_{x\in X}f'_\alpha(x)=0$ by Remark~\ref{rho},
and $f$ has totally balanced derivatives.
\qed

\begin{corollary}
\label{cor-no-bent}
If there is a $\psi\in\wh G$ such that
$(\C^X)_\psi=0$ (i.e. $(\wh X)_\psi=\emptyset$),
then there exists no bent function $f\in T^X$.
\end{corollary}

\pf In that case $|\wh{f_\psi}|=0$. \qed

\medskip
\begin{remark}\rm
The above corollary says that
 the condition ``$(\C^X)_\psi\ne 0$ for all $\psi\in\wh G$''
is a necessary condition for the existence of bent functions.

If the $G$-action on $X$ is not faithful, i.e.
the kernel $K$ of the action is nontrivial,
then there must be an irreducible character $\psi$ of $G$ which
takes nontrivial values on $K$, and hence $(\C^X)_\psi=0$,
cf. Remark~\ref{rem linear}(ii).
So by the above corollary, there exists no bent functions on $X$.

However, even if the $G$-action on $X$ is faithful,
there may still exist some $\psi\in\wh G$ such that
$(\C^X)_\psi=0$, and hence the bent functions on $X$ do not exist.
See Example~\ref{K4 no} below for such an example.
\end{remark}

Our next characterization of a bent function is given by its distance
from the set $(\C^X)_G$ of $G$-linear functions (cf. Definition~\ref{defn-linear}).
Recall that the distance between two subsets of $\C^X$ is defined
by Eqn~(\ref{dist sets}). The next theorem says that
the distance from a bent function to $(\C^X)_G$ is greater than the
distance from any non-bent unitary function to  $(\C^X)_G$. It also says that
$\sqrt{(m-1)n/m}$ is the best possible upper bound of the distance between
any unitary function and $(\C^X)_G$.

\begin{theorem}\label{bent max-dist}
Let $f \in T^X$. Then the following hold.

(i)~
$d(f, (\C^X)_G) \le \sqrt{\frac{(m-1)n}{m}}$.

(ii) $f$  is bent if and only if $d(f, (\C^X)_G) = \sqrt{\frac{(m-1)n}{m} }$.
\end{theorem}

\pf We have seen that
$$
(\C^X)_G=\bigcup_{\psi\in\wh G}(\C^X)_{\psi}.
$$
For any $G$-linear function $g$ there is a $\varphi\in\wh G$ such that
$g$ is $\varphi$-linear, i.e. $g=g_\varphi\in(\C^X)_\varphi$, and
$g_\psi=0$ for any $\psi \in \wh G$ such that $\psi\ne\varphi$.
Since any two different $G$-linear components are orthogonal
to each other, we can compute the distance between $f$ and $g$ as follows:
$$
{\rm d}(f,g)^2=|f-g|^2=\Big|\sum_{\psi\in\wh G}(f_\psi-g_\psi)\Big|^2
=|f_\varphi-g_\varphi|^2+\sum_{\psi\ne\varphi}|f_\psi|^2
\ge\sum_{\psi\ne\varphi}|f_\psi|^2;
$$
and the equality holds when $g=f_\varphi$.
By Corollary~\ref{cor13},
$$
\sum_{\psi\in\wh G}|f_\psi|^2=
\sum_{\psi\in\wh G} \langle f_\psi, f_\psi \rangle =
\left \langle \sum_{\psi\in\wh G} f_\psi, \sum_{\psi\in\wh G} f_\psi \right \rangle
= \langle f, f \rangle = |f|^2=n.
$$
So according to the definition of the distance in Eqn~(\ref{dist sets}), we have
$$
{\rm d}\big(f,(\C^X)_\varphi\big)^2=n-|f_\varphi|^2.
$$
Hence the square of the distance between $f$ and $(\C^X)_G$ is
$$
{\rm d}\big(f,(\C^X)_G\big)^2
=\min_{\varphi\in\wh G}\big\{n-|f_\varphi|^2\big\}
=n-\max_{\varphi\in\wh G}\big\{|f_\varphi|^2\big\}.
$$
By the equality $\sum_{\psi\in\wh G}|f_\psi|^2=n$ again, $| \wh G| = m$ implies that
$$
\max_{\varphi\in\wh G}\big\{|f_\varphi|^2\big\}\ge \frac{n}{m},
$$
where the equality holds if and only if
$|f_\psi|^2=|f_\varphi|^2$ for all $\psi,\varphi\in\wh G$.
In conclusion,
\begin{equation}\label{estim d}
{\rm d}\big(f,(\C^X)_G\big)^2\le n-\frac{n}{m}=\frac{(m-1)n}{m},
\end{equation}
and the equality in (\ref{estim d}) holds if and only if
$|f_\psi|^2=|f_\varphi|^2$ for all $\psi,\varphi\in\wh G$.
By Lemma \ref{lem-bent},
the equality in (\ref{estim d}) holds if and only if $f$ is bent.
\qed

By taking $X = G$ as the regular $G$-set, we have the next corollary from
Theorem \ref{bent deri-balance}, Theorem \ref{bent max-dist}
and Lemma \ref{lem-bent}. Note that the equivalence
of (i) and (ii) in Corollary \ref{cor-lsy} below was proved in \cite{LSY}.

\begin{corollary}
\label{cor-lsy}
Let $f: G \to T$ be a unitary function. Then the following are equivalent.

(i)~ $f$ is a bent function.

(ii)~ $f$ has totally balanced derivatives.

(iii)~ Among all functions in $T^G$, $f$ has the greatest distance
$\sqrt{|G|-1}$ from the set $(\C^G)_G$ of $G$-linear functions.

(iv)~ $| \langle f, \psi \rangle |$ are equal for all $\psi \in \wh G$.
\end{corollary}

\pf
The equivalence of (i), (ii), and (iii) is immediate from
Theorems \ref{bent deri-balance} and \ref{bent max-dist}.
Since $\wh G$ is a basis of $\C^G$, we may assume that $f =
\sum_{\psi \in \wh G} c_\psi \psi$, where $c_\psi \in \C$. Hence, the
$\psi$-component of $f$ is $f_\psi = c_\psi \psi$, for any $\psi \in \wh G$.
Thus,
$$
| \langle f, \psi \rangle | = | \langle c_\psi \psi, \psi \rangle | = |c_\psi|
= \sqrt{|\langle f_\psi, f_\psi \rangle |}, \quad \hbox{for any } \psi \in \wh G.
$$
So the equivalence of (i) and (iv) holds by Lemma \ref{lem-bent}. \qed

\begin{lemma}
\label{lem-f-x}
For any $f\in\C^X$ and $x\in X$, let $f_x\in\C^G$ be defined by
$f_x(\alpha)=f(\alpha x)$ for all $\alpha\in G$. Then
$\wh{f_x}(\psi)=mf_{\wb\psi}(x)$ for all $\psi\in\wh G$.
\end{lemma}

\pf It is a straightforward computation to see that
$$\wh{f_x}(\psi)=\sum_{\alpha\in G}f_x(\alpha)\psi(\alpha)
=\sum_{\alpha\in G}f(\alpha x)\wb\psi(\alpha^{-1})=(\wb\psi{*}f)(x)
=mf_{\wb\psi}(x).\eqno\qed$$

The next corollary is one of the main results of \cite{P09, PH},
where the $G$-bentness of $f\in T^X$ is defined by the condition (ii)
of  Corollary \ref{cor-bent}. So Corollary \ref{cor-bent} implies that
the $G$-bentness defined in \cite{P09, PH}
is equivalent to the bentness defined by Definition \ref{bent}.

\begin{corollary}
\label{cor-bent}
{\rm(Cf. \cite{P09, PH})}~
Let $f\in T^X$. Then the following
two statements are equivalent to each other:

(i)~ $f$ has totally balanced derivatives. That is, $f$ is a bent function by Definition \ref{bent}.

(ii)~ $\frac{1}{n}\sum_{x\in X}|\wh{f_x}(\psi)|^2=m$~ for all $\psi\in\wh G$.
That is, $f$ is a $G$-bent function by \cite[Definition 6]{P09}.
\noindent
\end{corollary}

\pf By Lemmas~\ref{lem-f-x} and \ref{<f,g>}, we have
$$
\frac{1}{n}\sum_{x\in X}|\wh{f_x}(\psi)|^2
=\frac{m^2}{n}\sum_{x\in X}|f_{\wb\psi}(x)|^2
=\frac{m^2}{n}\langle f_{\wb\psi},f_{\wb\psi}\rangle
=\frac{m^2}{n^2}\langle\wh{f_{\wb\psi}},\wh{f_{\wb\psi}}\rangle
=\frac{m^2}{n^2}|\wh{f_{\wb\psi}}|^2.
$$
Thus, (ii) holds if and only if
$|\wh{f_{\psi}}|^2=\frac{n^2}{m}$ for all $\psi\in\wh G$
if and only if $f$ has totally balanced derivatives by
Theorem \ref{bent deri-balance} and Lemma \ref{lem-bent}.
\qed

\begin{remark}\rm
For any $f, g \in \C^X$, the {\em pseudo-convolution} $f \boxtimes g$
of $f$ and $g$ is defined as (cf. \cite{PH})
$$
f \boxtimes g: \ G \to \C, \quad \alpha \mapsto \sum_{x \in X} \wb{f(x)}g(\alpha x).
$$
By Lemmas \ref{lem-or} and \ref{lem-hat-hat}, it is straightforward to show that
$$
\wh{(f \boxtimes g)}(\psi) = \frac{m}{n} \sum_{\lambda \in (\wh X)_\psi}
\wb{\wh{f}(\lambda)} \wh{g}(\lambda), \quad \hbox{for any}~ \psi \in \wh G.
$$
The equivalence of (i) and (ii) of Corollary~\ref{cor-bent}
can also be proved by the above equality.
\end{remark}

\section{Perfect nonlinear functions on $G$-sets}
\label{sect-perf}

As an application of the characterizations of bent functions on $G$-sets, in this section
we discuss the characterizations of perfect nonlinear functions from a $G$-set to an
abelian group. Our approach here is different from that of \cite{P09, PH}.
Let $X$ be a $G$-set as before, and let $H$ be an abelian group with
multiplicative operation.
Let $H^X$ denote the set of all functions from $X$ to $H$.
An $f\in H^X$ is said to be
{\em evenly-balanced} (cf. \cite{Xu1, Xu2}) if $|H|$ divides $|X|$ and
$$
\big | \{x \in X | f(x) = h \} \big | = \frac{|X|}{|H|}, \quad \hbox{for any}~
h \in H.
$$
An evenly-balanced function is also called a {\em balanced} or {\em uniformly
distributed} function in some literature. The {\em derivative} of
$f\in H^X$ in direction
$\alpha \in G$ is
$$
f'_\alpha: \ X \to H, \quad x \mapsto f(\alpha x)f(x)^{-1}.
$$

\begin{definition}\rm
(cf. \cite[Definition 1]{PH})
A function $f: X \to H$ is said to be {\em $G$-perfect nonlinear}
if for any $\alpha\in G\backslash\{1\}$, the function
$f'_\alpha$ is evenly-balanced.
\end{definition}

Any $g \in H^X$ induces a non-negative integral function $g^{\#}$ on $H$
as follows:
$$
g^{\#}: \ H \to \mathbb{N} \cup \{ 0 \}, \quad h \mapsto
\big | \{x \in X \,|\, f(x) = h \} \big |.
$$
Hence, $g^{\#}$ is constant on $H$ if and only if $g$ is evenly-balanced.
Thus, a function $f:X\to H$ is $G$-perfect nonlinear if and only if
for any $\alpha \in G$, $f^{\prime\#} _\alpha$ is constant on~$H$.

\begin{theorem}
\label{thm-nonlinear}
Let $f\in H^X$. Then following are equivalent.

(i)~ For any $\xi\in\wh H\backslash\{1\}$ the composition function
$\xi{\circ}f: X\to\C$ has totally balanced derivatives.

(ii)~ For any $\xi\in\wh H\backslash\{1\}$ the composition function
$\xi{\circ}f: X\to\C$ is bent.

(iii)~ The function $f:X\to H$ is $G$-perfect nonlinear.
\end{theorem}

\pf It is enough to show that (i) $\Leftrightarrow$ (iii).
Since $(\xi{\circ}f)(x)\in T$, we have
$(\xi{\circ}f)(x)^{-1}=\wb{(\xi{\circ}f)(x)}$, for any $x \in X$. So
\begin{eqnarray*}
\sum_{x\in X}(\xi{\circ}f)'_\alpha(x)
&=&\sum_{x\in X}(\xi{\circ}f)(\alpha x)\wb{(\xi{\circ}f)(x)}\\
&=&\sum_{x\in X}\xi\big(f(\alpha x)\big)\wb\xi\big(f(x)\big)
=\sum_{x\in X}\xi\big(f(\alpha x)\big)\xi\big(f(x)^{-1}\big)\\
&=&\sum_{x\in X}\xi\big(f(\alpha x)f(x)^{-1}\big)
=\sum_{x\in X}\xi\big(f'_\alpha(x)\big).
\end{eqnarray*}
For any $h\in H$, let $X(f'_\alpha,h)=\{x\in X\,|\,f'_\alpha(x)=h\}$.
Then $X$ is the disjoint union $X=\bigcup_{h\in H}X(f'_\alpha,h)$,
and the cardinality $|X(f'_\alpha,h)|=f^{\prime\#}_\alpha(h)$. So
\begin{equation}\label{xi f}
\sum_{x\in X}(\xi{\circ}f)'_\alpha(x)
=\sum_{h\in H}\sum_{x\in X(f'_\alpha,h)}\xi(h)
=\sum_{h\in H}f^{\prime\#}_\alpha(h)\xi(h)
=\wh{f^{\prime\#}_\alpha}(\xi).
\end{equation}
Thus, $(\xi{\circ}f)'_\alpha$ is balanced if and only if
$\wh{f^{\prime\#}_\alpha}(\xi)=0$.
Hence for any $\xi\in\wh H\backslash\{1\}$, the function
$(\xi{\circ}f)'_\alpha$  is balanced if and only if
$\wh{f^{\prime\#}_\alpha}$ is zero on $\wh H\backslash\{1\}$
 if and only if
$f^{\prime\#}_\alpha$ is constant on $H$ by Remark~\ref{rho}(ii).
That is, for any $\xi\in\wh H\backslash\{1\}$,
the function $\xi{\circ}f$ has totally balanced derivatives if and only if
$f$ is $G$-perfect nonlinear.
\qed

Taking $X = G$ to be the regular $G$-set, we have the next

\begin{corollary}\label{p-non-linear abelian}
{\rm (Cf. \cite{CD})}
Let $G, H$ be abelian groups, and $f: G \to H$ a function. Then the following are equivalent.

(i)~ $f$ is perfect nonlinear.

(ii)~ For any $\xi\in\wh H\backslash\{1\}$ the composition function
$\xi{\circ}f: G\to\C$ is bent.
\end{corollary}

Let $f \in H^X$. Then for any $x \in X$, there is a function (cf. \cite{P09, PH})
$$
f_x: \ G \to H, \quad \alpha \mapsto f(\alpha x).
$$
Also for any $\xi \in \wh H$, there is a function
$(\xi {\circ}f)_x: \, G \to T, \,
\alpha \mapsto (\xi {\circ}f)(\alpha x)$. Note that $(\xi {\circ}f)_x = \xi {\circ}f_x$, for any
$x \in X$. The next corollary is immediate from Theorem
\ref{thm-nonlinear} and Corollary \ref{cor-bent}.

\begin{corollary}
{\rm (cf. \cite[Theorems 5 and 7]{P09})}
Let $f \in H^X$. Then the following are equivalent.

(i)~ $f$ is $G$-perfect nonlinear.

(ii)~ For any $\xi \in \wh H\backslash\{1\}$ and $\alpha \in G$,
$$
\frac{1}{|X|} \sum_{x \in X} \left | \wh{(\xi {\circ}f_x)}(\alpha) \right|^2 = |G|.
$$
\end{corollary}

\section{Examples}
\label{sect-ex}

In this section we present a few examples that explain the theory developed
in the previous sections.

\begin{example}\rm Assume that $X=G$ is the regular $G$-set.
As mentioned in Remark~\ref{rem standard},
the \gdual\ $\wh X$ is unique up to rescaling by $T$, and
the typical choice of $\wh X$ is just the dual group~$\wh G$.
So the theory developed in previous sections includes the corresponding theory
for finite abelian groups as a special case. For example,  
 some well-known results in \cite{CD,LSY,S02} as well as other properties 
of bent functions on finite abelian groups are given in
Corollary \ref{cor-lsy} and Corollary \ref{p-non-linear abelian}
 as immediate consequences.  
\end{example}

The next example gives a $G$-set on which there exists no bent function.

\begin{example}\label{K4 no}\rm
Let $G=\{1,\alpha,\beta,\gamma\}$ be the Klein four group. That is, $G$ is
an abelian group such that
$$
\alpha^2=\beta^2=\gamma^2=1,~
            \alpha\beta=\gamma,~ \beta\gamma=\alpha,~
            \gamma\alpha=\beta.
$$
Then $\wh G=\{\psi_1=1,\psi_2,\psi_3,\psi_4\}$
is given by Table 6.1.
\begin{table}[h]\label{character}
\begin{center}
$\begin{array}{|c|c c c c|}\hline
 & 1 & \alpha & \beta & \gamma \\ \hline
\psi_1 & 1 & 1 & 1 & 1\\ \psi_2 & 1 & 1 &-1&-1\\
\psi_3 & 1 & -1 & 1 & -1 \\ \psi_4 & 1 & -1 & -1 & 1\\\hline
\end{array}$
\caption{Character Table of the Klein Four Group}\end{center}
\end{table}

Let $X=\{x_1,x_2,x_3,x_4\}$ be a faithful $G$-set
with two orbits $X_1$ and $X_2$ as follows:

$\bullet$~ $X_1=\{x_1,x_2\}$,~
$1$ and $\alpha$ fix both points $x_1$ and $x_2$,
while $\beta$ and $\gamma$ interchange the two points;

$\bullet$~ $X_2=\{x_3,x_4\}$,~
$1$ and $\beta$ fix both points $x_3$ and $x_4$,
while $\gamma$ and $\alpha$ interchange the two points.

\noindent
We can take $\wh X=
\{\lambda_1,\lambda_2,\lambda_3,\lambda_4\}$
as in Table~\ref{t2} (to simplify the table,
we list $\frac{1}{\sqrt 2}\lambda_i$ instead of $\lambda_i$).
\begin{table}[h]\label{t2}
\begin{center}
$\begin{array}{|c|c c c c|}\hline
 & x_1 & x_2 & x_3 & x_4 \\
 \hline
\frac{1}{\sqrt 2}\lambda_1 & 1 & 1 & 0 & 0\\
\frac{1}{\sqrt 2}\lambda_2 & 1 & -1 & 0 & 0\\
\frac{1}{\sqrt 2}\lambda_3 & 0 & 0 & 1 & 1 \\
\frac{1}{\sqrt 2}\lambda_4 & 0 & 0 & 1 & -1  \\ \hline
\end{array}$
\caption{The \gdual\ in Example~\ref{K4 no}}\end{center}
\end{table}

Hence, the $G$-linear components are
$$
(\wh X)_{\psi_1}=\{\lambda_1,\lambda_3\},~
(\wh X)_{\psi_2}=\{\lambda_2\},~
(\wh X)_{\psi_3}=\{\lambda_4\},~
(\wh X)_{\psi_4}=\emptyset.
$$
Since one of the $G$-linear components is empty, there exists no
bent function $f\in T^X$ by Corollary \ref{cor-no-bent}.
\end{example}

The next example gives a $G$-set $X$ and a bent function on $X$.

\begin{example}\label{K4 yes}\rm
As above in Example~\ref{K4 no},
let $G=\{1,\alpha,\beta,\gamma\}$ be the Klein four group
and $\wh G=\{\psi_1,\psi_2,\psi_3,\psi_4\}$.
But this time we consider the $G$-set
$X=\{x_1,x_2,x_3,x_4,x_5,x_6\}$ with three orbits:

$\bullet$~ $X_1=\{x_1,x_2\}$,~
$1$ and $\alpha$ fix both points $x_1$ and $x_2$,
while $\beta$ and $\gamma$ interchange the two points;

$\bullet$~ $X_2=\{x_3,x_4\}$,~
$1$ and $\beta$ fix both points $x_3$ and $x_4$,
while $\gamma$ and $\alpha$ interchange the two points;

$\bullet$~ $X_3=\{x_5,x_6\}$,~
$1$ and $\gamma$ fix both points $x_5$ and $x_6$,
while $\alpha$ and $\beta$ interchange the two points.

\noindent
We can take $\wh X=
\{\lambda_1,\lambda_2,\lambda_3,\lambda_4,\lambda_5,\lambda_6\}$
as in Table~\ref{t3} (to simplify the table,
we list $\frac{1}{\sqrt 3}\lambda_i$ instead of $\lambda_i$).
\begin{table}[h]\label{t3}
\begin{center}
$\begin{array}{|c|c c c c c c|}\hline
 & x_1 & x_2 & x_3 & x_4 & x_5 & x_6 \\
 \hline
\frac{1}{\sqrt 3}\lambda_1 & 1 & 1 & 0 & 0 & 0 & 0\\
\frac{1}{\sqrt 3}\lambda_2 & 1 & -1 & 0 & 0 & 0 & 0\\
\frac{1}{\sqrt 3}\lambda_3 & 0 & 0 & 1 & 1 & 0 & 0\\
\frac{1}{\sqrt 3}\lambda_4 & 0 & 0 & 1 & -1 & 0 & 0\\
\frac{1}{\sqrt 3}\lambda_5 & 0 & 0 & 0 & 0 & 1 & 1\\
\frac{1}{\sqrt 3}\lambda_6 & 0 & 0 & 0 & 0 & 1 & -1 \\ \hline
\end{array}$
\caption{The \gdual\ in Example~\ref{K4 yes}}\end{center}
\end{table}

We can check that the $G$-linear components of $\C^X$ are
$$
(\wh X)_{\psi_1}=\{\lambda_1,\lambda_3,\lambda_5\},~
(\wh X)_{\psi_2}=\{\lambda_2\},~
(\wh X)_{\psi_3}=\{\lambda_4\},~
(\wh X)_{\psi_4}=\{\lambda_6\}.
$$
Let $\omega=\frac{-1+\sqrt{-3}}{2}$ be a primitive third root of unity.
Take $f\in T^X$ as follows:
$$ f(x_j)=\omega^{(1+(-1)^j)/2}
  =\begin{cases}1, & j=1,3,5;\\\omega, & j=2,4,6. \end{cases}
$$
Then
$$
 \sum_{x\in X_j}f'_\alpha(x)
 =\sum_{x\in X_j}f(\alpha x)f(x)^{-1}
 =\begin{cases} 1+1=2, & j=1;\\
  1\cdot\omega^{-1}+\omega\cdot 1=-1, & j=2,3. \end{cases}
$$
So $\sum\limits_{x\in X}f'_\alpha(x)=0$. Similarly,
$\sum\limits_{x\in X}f'_\beta(x)=\sum\limits_{x\in X}f'_\gamma(x)=0$.
That is, $f$ has totally balanced derivatives.

On the other hand,
\begin{eqnarray*}
\langle\wh f_{\psi_1},\wh f_{\psi_1}\rangle & = &
\sum_{\lambda\in(\wh X)_{\psi_1}}|\wh f(\lambda)|^2
=\sum_{j=1,3,5}\big|\sum_{x\in X}f(x)\lambda_j(x)\big|^2 \\
 & = & \sum_{j=1,3,5}|1+\omega|^2=3|1+\omega|^2 = 3,
\end{eqnarray*}
$$
\langle\wh f_{\psi_2},\wh f_{\psi_2}\rangle=
\sum_{\lambda\in(\wh X)_{\psi_2}}|\wh f(\lambda)|^2
=\big|\sum_{x\in X}f(x)\lambda_2(x)\big|^2
=|1-\omega|^2 = 3.
$$
Similarly,
$\langle\wh f_{\psi_3},\wh f_{\psi_3}\rangle=
 \langle\wh f_{\psi_4},\wh f_{\psi_4}\rangle=|1-\omega|^2 = 3$.
In conclusion, we have
$\langle\wh f_{\psi},\wh f_{\psi}\rangle=3$,
$\forall$~$\psi\in\wh G$, and $f$ is a bent function.
\end{example}

Finally, we give an example of a $G$-perfect nonlinear function.

\begin{example}\rm We continue Example~\ref{K4 yes} and further
take $H=\{1,h,h^2\}$ with $h^3=1$ to be a cyclic group of order $3$.
Let $g:X\to H$ be as follows:
$$ g(x_j)=h^{(1+(-1)^j)/2}
  =\begin{cases}1, & j=1,3,5;\\ h, & j=2,4,6. \end{cases}
$$
It is known that
$\wh H=\{1,\xi,\xi^2\}$, where $\xi(h^i)=\omega^i$, $i=0,1,2$.
Then the composition function $\xi{\circ}g:X\to\C$ is just the
function $f$ in Example~\ref{K4 yes}, and hence
$\xi{\circ}g$ is a bent function on $X$. Similarly
we can check that $\xi^2{\circ}g$ is also a bent function on $X$.
So $g:X\to H$ is a $G$-perfect nonlinear function from the $G$-set
$X$ to the abelian group $H$.
In fact, one can check directly that
$$
 g'_\alpha(x_j)=g(\alpha x_j)g(x_j)^{-1}
 =\begin{cases}1,& j=1,2;\\ h, & j=3,5;\\ h^2, & i=4,6. \end{cases}
$$
Hence, $g^{\prime\#}_\alpha(h^i)=2$ for $i=0,1,2$.
Similarly, $g^{\prime\#}_\beta=g^{\prime\#}_\gamma=2$ are
constant functions on $H$, too.
\end{example}

\section*{Acknowledgments}

This work was done while the first author
was visiting the second author at Eastern Kentucky University
in Spring 2014; he is grateful for the hospitality.
The work of the first author is supported by NSFC
with grant numbers 11171194 and 11271005.

\end{document}